\documentclass[a4paper,11pt]{article}
\usepackage{pos}
\usepackage{wrapfig}

\title{Going Big for Phase III of the Project~8 Neutrino Mass Experiment}

\author*[a]{Juliana Stachurska}

\affiliation[a]{Laboratory for Nuclear Science, Massachusetts Institute of Technology\\
  Cambridge, Massachusetts 02139, USA}

\emailAdd{jstach@mit.edu}
\onbehalf{on behalf of the Project~8 collaboration}

\abstract{Project 8 is a next generation experiment aiming to directly measure the neutrino mass using the tritium endpoint method with a targeted sensitivity of 40~meV. Having established a new measuring technique, Cyclotron Radiation Emission Spectroscopy (CRES), the next development phase will demonstrate CRES on a large source volume, culminating in a pilot-scale CRES experiment with atomic tritium. A promising option is a mode-filtered, cylindrical, resonant cavity in which cyclotron radiation from magnetically trapped beta electrons only couples to the lowest eigenmode, maximizing effective volume and minimizing signal complexity. 
Recent progress in the experiment design, including a small scale cavity CRES proof-of-concept apparatus to demonstrate CRES in cavities and its scalability to large volumes are described.}

\FullConference{XVIII International Conference on Topics in Astroparticle and Underground Physics (TAUP2023)\\
 28.08-01.09.2023\\
University of Vienna\\}

\begin{document}
\maketitle

\section{Direct Neutrino Mass Measurement with Project~8}
The unknown absolute neutrino mass scale is one of the outstanding puzzles of both particle physics and cosmology. Laboratory measurements of the tritium beta decay spectrum endpoint region offer a direct and model-independent way to measure the electron-weighted effective neutrino mass $m_{\beta}$, defined as $m_{\beta}^2 = \sum_i |U_{ei}|^2 m_i^2$. 
Project~8 is a proposed experiment using the tritium endpoint method and targeting a sensitivity of $40 \mathrm{\,meV}$, covering the entire mass range allowed in the inverted mass ordering. Project~8 established Cyclotron Radiation Emission Spectroscopy (CRES) \cite{Monreal_2009}, a non-destructive technique to measure electron energies \cite{Project8:PRL2015} via
\begin{equation}
	f_c = \frac{1}{2 \pi} \frac{e \langle B \rangle}{m_e + E/c^2},
	\label{eq:CRES}
\end{equation}
where $f_c$ is the frequency of cyclotron radiation emitted by electrons with energy $E$ in a magnetic field $B$.
Recently, the first CRES-based neutrino mass upper limit was extracted from the tritium decay spectrum~\cite{Project8:2023PRL}.
The next phase will establish atomic tritium production, cooling and trapping -- see \cite{TAUPLarisa} on recent progress on the atomic source --, and scale CRES to large volumes. 

\section{A Cavity-Based CRES Experiment}
\setlength{\columnsep}{20pt}
\begin{wrapfigure}{l}{.5\textwidth}
	\vspace{-7mm}
	\includegraphics[width=0.5\textwidth]{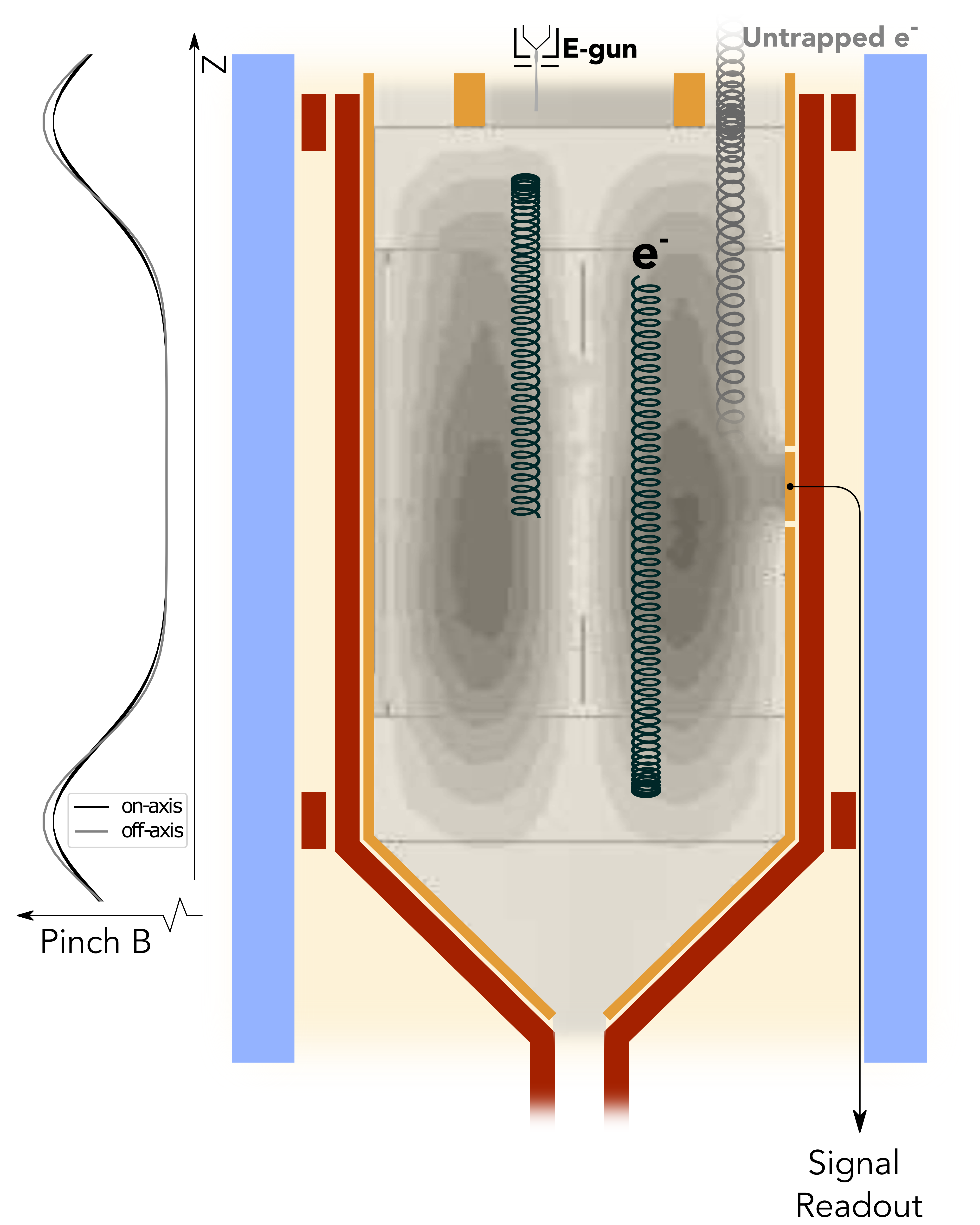}
	\vspace{-7mm}
	\caption{Project~8 cavity concept with cavity (orange), multipole magnet for atom trapping and pinch coils for electron trapping (red) and solenoid magnet to induce cyclotron motion (blue). Electrons couple to the  TE$_{011}$ mode (grey), read out from the center. The electron trapping field is shown on the left.}
	\vspace{-5mm}
	\label{fig:CavityConcept}
\end{wrapfigure}
The initial phases of Project~8 detected CRES in a magnetic field with strength of $\sim1$~T, corresponding to a cyclotron frequency of $\sim 26$~GHz for electron energies of $\sim 18.6$~keV, close to the tritium endpoint, using a WR-42 waveguide to guide the RF signal from the gas cell and the electrons to the amplifier. The final Project~8 experiment will monitor a large quantity of electrons released from atomic tritium beta decay, and thus calls for a large detector volume. The low-field seeking hyperfine states of atomic tritium will be confined by a combination of magnetic field walls and gravity. Dipolar spin-flip leads to losses from the trap and requires CRES fields much lower than $1$~T. However, the Larmor power of the emitted cyclotron radiation depends on the frequency as $P \propto f_c^2$. Cavities naturally provide large volumes at low resonant frequencies, while the Purcell enhancement \cite{Purcell} can compensate for the lower radiated powers. 
The final Project~8 experiment is currently being designed to use cavities with resonant frequency of $325$~MHz as the detector volume. The design needs to be compatible with atomic tritium operations, such as injecting a beam of atomic tritium into it, as well as the introduction of calibration electrons. The atomic tritium will be confined by a combination of magnetic fields and gravity, necessitating a tall structure and a closely fitting magnet array around the cavity. 
The chosen axial field of strength $11$~mT corresponds to an electron cyclotron frequency of $f_c = 325$~MHz at the tritium endpoint. The electron is trapped axially by higher magnetic field walls from pinch coils. Figure \ref{fig:CavityConcept} shows the concept of a cavity CRES experiment. The mode used is a TE$_{011}$ mode, to which the readout is designed to achieve an appropriate loaded $Q$, or, a resonance bandwidth covering the electron energy and pitch angle range of interest. To maximize volume at the given frequency, the cavity is 11~m tall at a diameter of 1.1~m. 

\section{The Cavity CRES Apparatus}
\begin{figure}[tbh]
	\centering
	\includegraphics[width=0.8\textwidth]{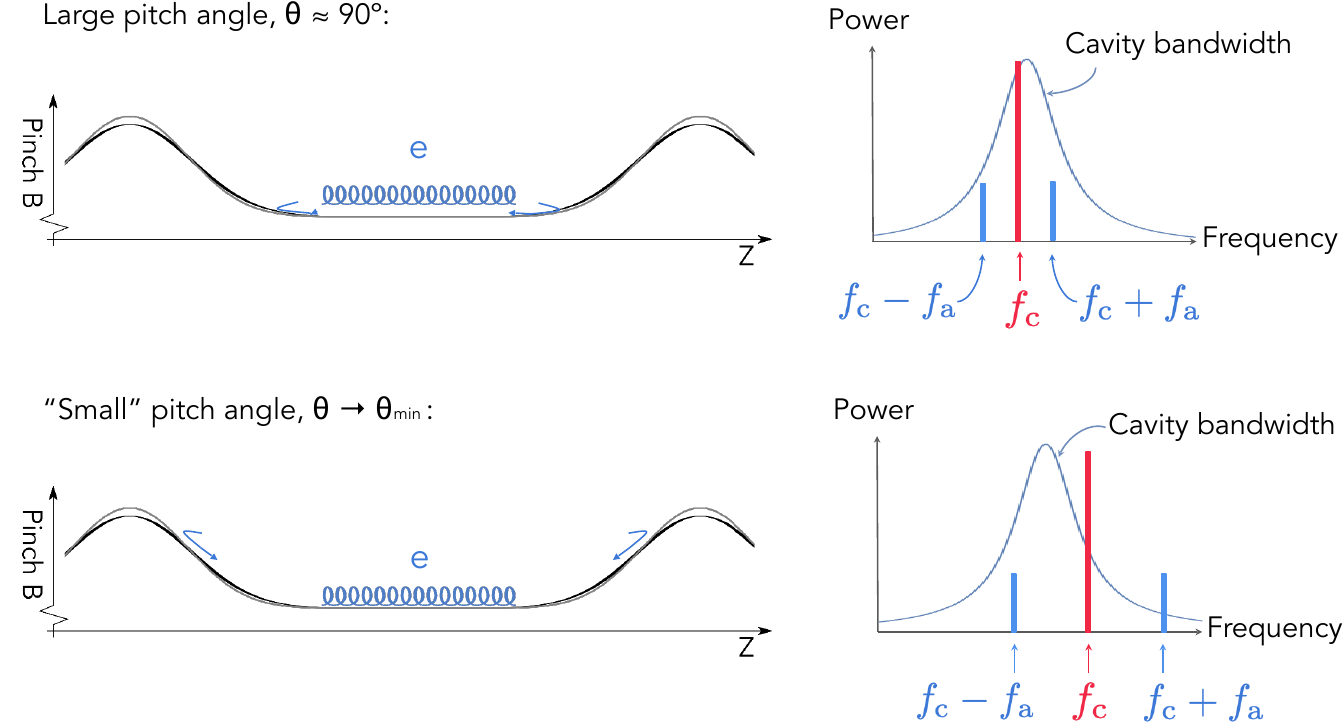}
	\caption{Motion of axially trapped electrons in the cavity for large and small pitch angles, and the resulting signal structure. Electrons with smaller pitch angles will experience a higher magnetic field, leading to a higher cyclotron frequency. The effect can be corrected by detecting the lower sideband, a result of the pitch-angle dependent axial motion.}
	\label{fig:bandwidth}
\end{figure}
The first demonstration of the cavity-CRES viability will be at a field of $1$~T and a cyclotron frequency of $26$~GHz. This allows for the re-use of RF equipment, software, as well as a $1$~T MRI magnet with excellent field uniformity for better time and cost efficiency. The cavity will be read out with a WR-42 waveguide. Two sources of CRES electrons will be used: an electron gun, with electrons Mott-scattering inside the cavity on Helium gas, and conversion electrons from $^{83m}$Kr. The design of the cavity is open-ended, and has the length-to-diameter ratio of $L/D=10$ of the final cavity concept. 
The cavity CRES apparatus will establish cavities as CRES detectors, verifying simulated signal structure and volume and pitch angle acceptance. 
To achieve the desired energy resolution of $0.3$~eV, individual electrons' experienced magnetic fields need to be known with high precision. As noted in Equation \ref{eq:CRES}, the cyclotron frequency is a function of the average magnetic field the electron experiences. Electrons with high pitch angles $\theta \approx 90^{\circ}$ have momenta almost orthogonal to the axial trapping field. They will be able to move only in regions where the field is close to its lowest value and experience $\langle B \rangle \approx B_{\mathrm{min}}$. Electrons with lower pitch angles will be able to climb higher up the magnetic trapping walls, leading to $\langle B \rangle > B_{\mathrm{min}}$, and thus to a higher $f_c$. However, the axial frequency $f_a$ depends on the pitch angle and the shape of the trapping field. 
The sidebands appearing at $f_c \pm n f_a, \, n=1,2,3,...$ can be used to correct for the higher average magnetic field experienced. As $f_c$ increases with decreasing pitch angle, we will use the lower sidebands for the field correction, which will stay within the cavity bandwidth. The detection and reconstruction concept is illustrated in Figure \ref{fig:bandwidth}. The demonstrator is designed to trap and reconstruct electrons with pitch angles $\geq 88^{\circ}$. In addition to establishing the electron gun as a calibration device, $^{83m}$Kr conversion electron lines will be measured.

\section{Low-frequency CRES Detection}
Development of the CRES technique to the lower magnetic fields needed for atomic tritium trapping poses its own set of challenges due to the low collected powers and the field homogeneity and stability requirements over large volumes. 
We are designing a demonstrator for low-frequency and low-power CRES detection around $1$~GHz, where the collected power will be on the order of $10^{-18}$~W. 
While $1$~T magnets providing homogenous fields over a significant volume are readily available commercially, this is not true at much lower fields. The solenoid magnet providing the CRES field at $\sim 35$~mT will be custom-designed. 
With the successful operation of the low-frequency CRES demonstrator, cavities will be established as large-volume CRES detectors. This must include the demonstration of high resolution at low power, necessitating the creation of a highly uniform, stable low magnetic field with a custom magnet.
Following the demonstrator's success, the Phase-III pilot atomic tritium experiment can be built with a cavity detector at a resonant frequency of $325$~MHz, providing an $11$~m$^3$ volume for low-density atomic tritium gas. At a projected sensitivity of $m_{\beta}<100 \mathrm{\,meV}$ after one year of data taking, the pilot atomic tritium neutrino mass experiment will be world-leading. 

\bibliographystyle{JHEP}
\bibliography{proceedings}
\end{document}